\documentclass[doublecol]{epl2}
\usepackage[T1]{fontenc}
\usepackage[latin1]{inputenc}
\usepackage{amsmath}
\usepackage{graphicx}

\title{Intrinsic nanoscale inhomogeneity in ordering systems due to
elastic-mediated interactions}

\shorttitle{Intrinsic nanoscale inhomogeneity in ordering systems}

\author{I. K. Razumov\inst{1,2} \and Yu. N. Gornostyrev\inst{1,2} \and M. I. Katsnelson\inst{3}}
\shortauthor{I. K. Razumov \etal}

\institute{
  \inst{1} Institute of Metal Physics, 620219 Ekaterinburg,
Russia \\
  \inst{2} Institute of Quantum Materials Science, 620107
Ekaterinburg, Russia \\
  \inst{3} Institute for Molecules and Materials, Radboud University Nijmegen,
6525 ED Nijmegen, The Netherlands
 }




\abstract{Phase diagram and pattern formation in two-dimensional Ising
model with coupling between order parameter and lattice vibrations is
investigated by Monte-Carlo simulations. It is shown that if the
coupling is strong enough (or phonons are soft enough) a
short-range order exists in disordered phase for a broader
temperature interval. Different types of this short-range order
(stripe-like, checkboard-like, etc.) depending on the temperature
and model parameters are investigated. With further increase of the coupling,
a reconstruction of the ground state happens and new ordered phases appear
at low enough temperatures.}

\begin{document}

\maketitle

\section{Introduction}

Traditional views assume that under the thermodynamic equilibrium
conditions a system should either be homogeneous or consist of
macroscopically large domains of homogeneous phases. It appears
frequently, however, that equilibrium or very long-lived
metastable states occur with a nanoscale or mesoscale
heterogeneity (for a general review of these phenomena in various
physical and chemical systems see, e.g.,
Refs.\cite{seul,chem,dagotto,JPCMreview}). It is commonly accepted
now that the formation of the mesoscale heterogeneity is a result
of frustrations in the system which can result from either
geometric factors~\cite{JPCMreview,kleman,nelson,pack} or
competing interactions, the long-ranged forces such as Coulomb or
dipole-dipole interactions being of primary
importance~\cite{JPCMreview,emery,kivelson,nussinov,debell,schmalian,jagla,EPL}.

``Heterophase'' fluctuations in metallic alloys~\cite{krivoglaz}
provide us interesting examples of intrinsic nanoscale
inhomogeneities. In such cases, a system behaves like an ensemble
of nanosize particles of one phase embedded into the matrix, e.g.,
so-called athermic $\omega$-phase in bcc host which is observed in
some Ti- and Zr-based alloys~\cite{collings}, as well as in
Cr$_{1-x}$Al$_{x}$~\cite{sudareva}. This peculiar structural state
leads to strong anomalies of electronic properties~\cite{KT} and
ultrasound attenuation~\cite{GKT}. The ``athermic $\omega$-phase''
is a term to describe rather some short-range order in atomic
positions than real phase since it never exists in the bulk
(atomic positions in this ``phase'' are intermediate between bcc
structure and true $\omega$-phase existing in Ti and Zr under
pressure). It would be not surprising to observe strong
short-range order in the close vicinity of a second-order phase
transition. On the contrary, these heterophase fluctuations
sometimes arise in a broad temperature and concentration domain.
Nature of this state is still unknown (see discussions in
Refs.\cite{pack,krivoglaz}).

A well-pronounced short-range order or nanoscale inhomogeneity in
a broad temperature interval are often observed in magnetic alloys
with a strong coupling between magnetism and lattice (or chemical
composition), such as Cu-Mn alloys~\cite{udov,tsunoda} or Fe-Ni
Invar alloys~\cite{menshikov,book}. As mentioned above it is a tendency
now to connect intrinsic inhomogeneities in various systems with
long-range interactions. One may expect therefore that a
long-range character of elastic deformations is solids could be
relevant for the problem under discussion.

To clarify a role of elastic-mediated interactions in possible
pattern formation we have investigated the simplest model of Ising
order parameter coupled with phonons at the square lattice. Basing
on the results of Monte Carlo simulations for this model we show
that, indeed, this coupling can result in a formation of various
nanosize-scale structures.

\section{Model and computational details}

The two-dimensional Ising model~\cite{onsager,onsager1} is a
prototype, exactly solvable model of order-disorder phase
transitions in magnetic systems, ordering alloys, etc. To be
specific we will use further terms ``magnetic'' and ``spins'' to
describe the ordering phenomena under consideration. To describe
the effects of magnetoelastic (spin-lattice) interactions, we
proceed with the Hamiltonian
\begin{equation}
H = \frac K2 \sum_{i,j}\Delta_{ij} ({\bf u}_i-{\bf u}_j)^2 +
\frac 12 \sum_{i,j} J_{ij}S_iS_j \label{H1}
\end{equation}
where $\Delta_{ij}=1$ if $i,j$ are the nearest neighbors and is
equal to zero, otherwise, ${\bf u}_i$ are atomic displacement
vectors and spin variables $S_i = \pm 1$. The spin--lattice
coupling is taken into account via coordinate dependence of the
exchange parameters $J_{ij}=J({\bf R}_i+{\bf u}_i -{\bf R}_j-{\bf
u}_j)$, ${\bf R}_i$ being square lattice vectors. Assuming that
the displacements are small the exchange parameters can be written
in the linear approximation
\begin{equation}
J_{ij}= J_{ij}^0 + J_{ij}^{\prime} {\bf n}_{ij}({\bf u}_i -{\bf
u}_j),\label{Jij}
\end{equation}
where ${\bf n}_{ij}$ is the unit vector in direction of ${\bf
R}_i-{\bf R}_j$.

Substituting Eq.(\ref{Jij}) into Eq.(\ref{H1}) and assuming
periodic boundary conditions we obtain
\begin{equation}
H = \frac 12 \sum_{i,j} \Phi _{ij}^{\alpha \beta} u_i^{\alpha}
u_j^{\beta}
    + \frac 12 \sum_{i,j} J_{ij}^{0} S_iS_j +  \sum_{i} P_i^{\alpha}
    u_i^{\alpha}.
\label{H2}
\end{equation}
where $\Phi _{ij}^{\alpha \beta} = K n_{ij}^{\alpha}n_{ij}^{\beta}
\Delta_{ij}$. The dependence of the exchange parameters on
interatomic distances results in the last term in right-hand-side
of Eq.(\ref{H2})
\begin{equation}
P_i^{\alpha} = \sum_{j} n_{ij}^{\alpha} J_{ij}^{\prime} S_iS_j
\label{P}
\end{equation}
describing forces acting on atoms due to spin redistribution.
These forces initiate the displacements of atoms into new
equilibrium positions ${\bf R}+{\bf u}_0$ determined by the
expression
\begin{equation}
u_0^{\alpha}({\bf R}_i) = \sum_j G_{\alpha \beta}({\bf R}_{ij})P_j^{\beta}
\label{u0}
\end{equation}
where ${\bf R}_{ij} \equiv {\bf R}_i - {\bf R}_j$, $G_{\alpha
\beta}({\bf R}_{ij})$ is the static lattice Green's function
determined by the equation
\begin{equation}
 \sum_j \Phi_{ij}^{\alpha \beta} G_{\alpha \beta}({\bf R}_{ij}) =
-\delta_{ij}\delta_{\alpha \beta}. \label{GF}
\end{equation}
Replacing the variables ${\bf u}={\bf u}_0 + {\bf g}$ one can
represent the partition function $Z$ as
\begin{eqnarray}
&& Z  \sim  \int d{\bf g}_1...d{\bf g}_N \exp
\left(- \frac {\beta}2 \sum_{i,j}
\Phi _{ij}^{\alpha \beta} g_i^{\alpha} g_j^{\beta} \right) \nonumber\\
&& \times \sum_{S_i} \exp (-\beta H_{eff}) \label{Z2}
\end{eqnarray}
where the effective spin Hamiltonian
\begin{equation}
H_{eff}= \frac 12 \sum_{i,j} J_{ij}^{0} S_iS_j+ \frac 12 \sum_{ij}
P_i^{\alpha}G_{\alpha \beta}({\bf R}_{ij}) P_j^{\beta}
\label{Veff}
\end{equation}
and $\beta=1/T$ is the inverse temperature.

Due to harmonic approximation for the potential energy of atomic
displacements and linear approximation for the magnetoelastic
coupling the phonon and spin subsystems turn out to be totally
separated after the change of variables. The statistical
properties of the system is completely determined by the
Hamiltonian $H_{eff}$ which depends on spin variables only. The
last term in the right-hand-side of Eq.(\ref{Veff}) reads
\begin{equation}
\frac 18 \sum_{ij} \sum_{kl}  n_{ik}^{\alpha} J_{ik}^{\prime} S_iS_k
G_{\alpha \beta}({\bf R}_{ij}) n_{jl}^{\beta} J_{jl}^{\prime} S_jS_l
\label{interac}
\end{equation}
and describes indirect long-range spin-spin interactions via
lattice distortions.

To investigate equilibrium properties of the spin system with the
Hamiltonian (\ref{Veff}) we have carried out the Monte Carlo
simulations for the square lattice with 100 $\times$ 100 sites and
periodic boundary conditions, using a standard Metropolis
algorithm~\cite{MC}. To provide the Gibbs distribution $P \sim
\exp (-\beta H_{eff})$ we have performed up to 10$^4$ flips for each
lattice spin.

We are interested in the two-dimensional spin system embedded into
three-dimensional elastic medium. For two-dimensional continuum
the Green's function is pathological, with logarithmic growth at
large distances. Instead, we use the expression
\begin{equation}
G_{\alpha \beta} = \frac {\delta_{\alpha \beta}} {4\pi \mu} \frac
{Si (2\pi R/a)}{R} \label{GF}
\end{equation}
valid in the framework of quasi-continuum approach~\cite{Kunin}
with isotropic Debye model for the phonon spectra. Here $Si(x)$ is
the integral sine function~\cite{abram}, $a$ is the lattice
parameter, $\mu$ is the shear modulus which can be
expressed~\cite{Kunin} in terms of the force constants
$\Phi^{\alpha \beta}$. The Green's function (\ref{GF}) has
asymptotic  behavior $ G_{\alpha \beta}(R) \propto 1/R$ at large
distances. To speed up computations, we use its truncation by
multiplying expression (\ref{GF}) by $\exp (-(R/L)^4)$. The length
$L$ and phonon-induced interaction cut-off radius were chosen at
5th and 13th neighbors, respectively; these values ensure the
convergency and stability of computational results.

For the bare exchange parameters $J^{0}_{ij}$ the nearest-neighbor
({\it nn}) interactions have been taken into account. The choice of
{\it nn} parameter $|J^{0}_1| =1$ determines the energy units.

\begin{figure*}[!htb]
\begin{tabular}{cccc}
{\bf a} & {\bf b} & {\bf c} & {\bf d} \\
\includegraphics[width=4.00cm,angle=0,clip]{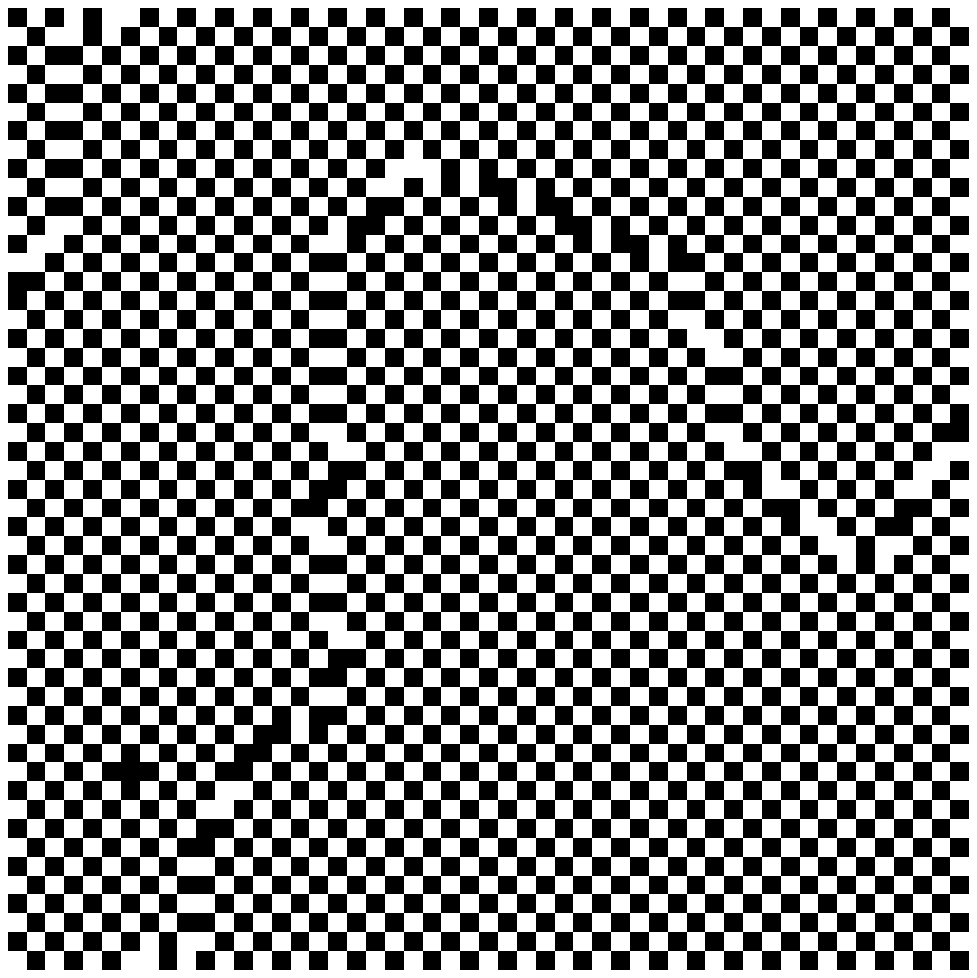}
&
\includegraphics[width=4.00cm,angle=0,clip]{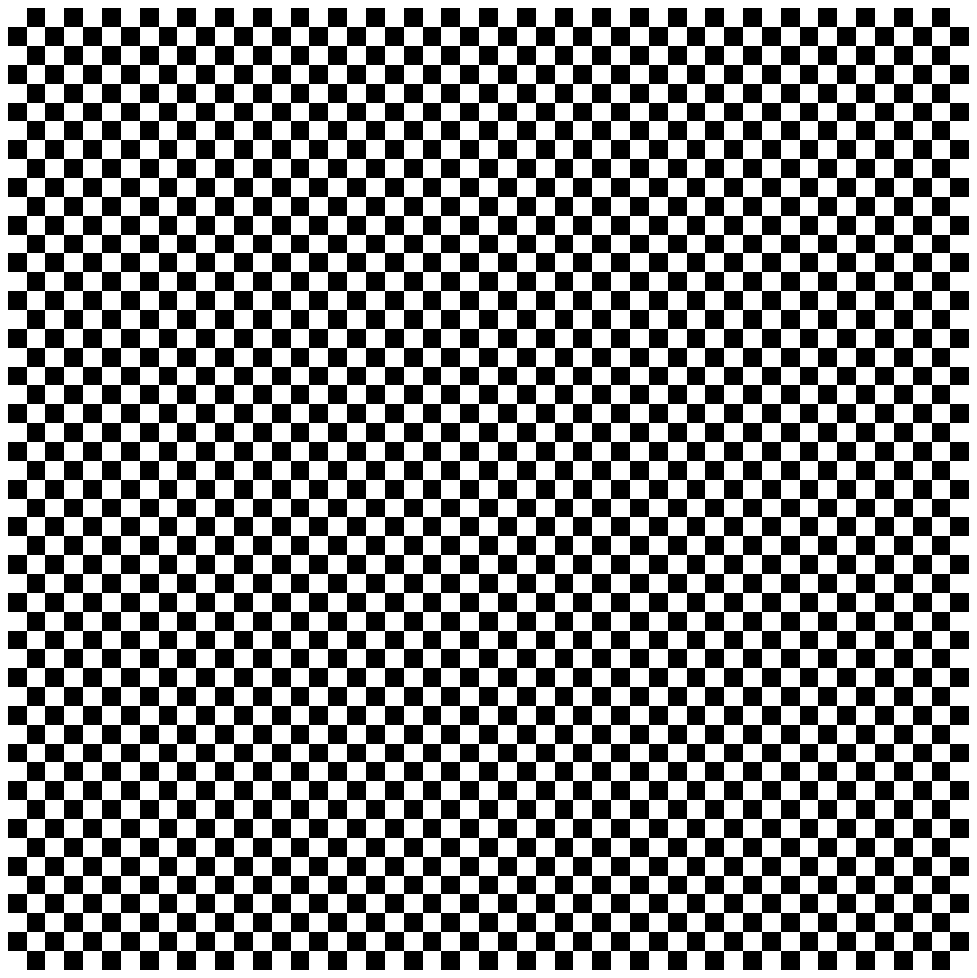}
&
\includegraphics[width=4.00cm,angle=0,clip]{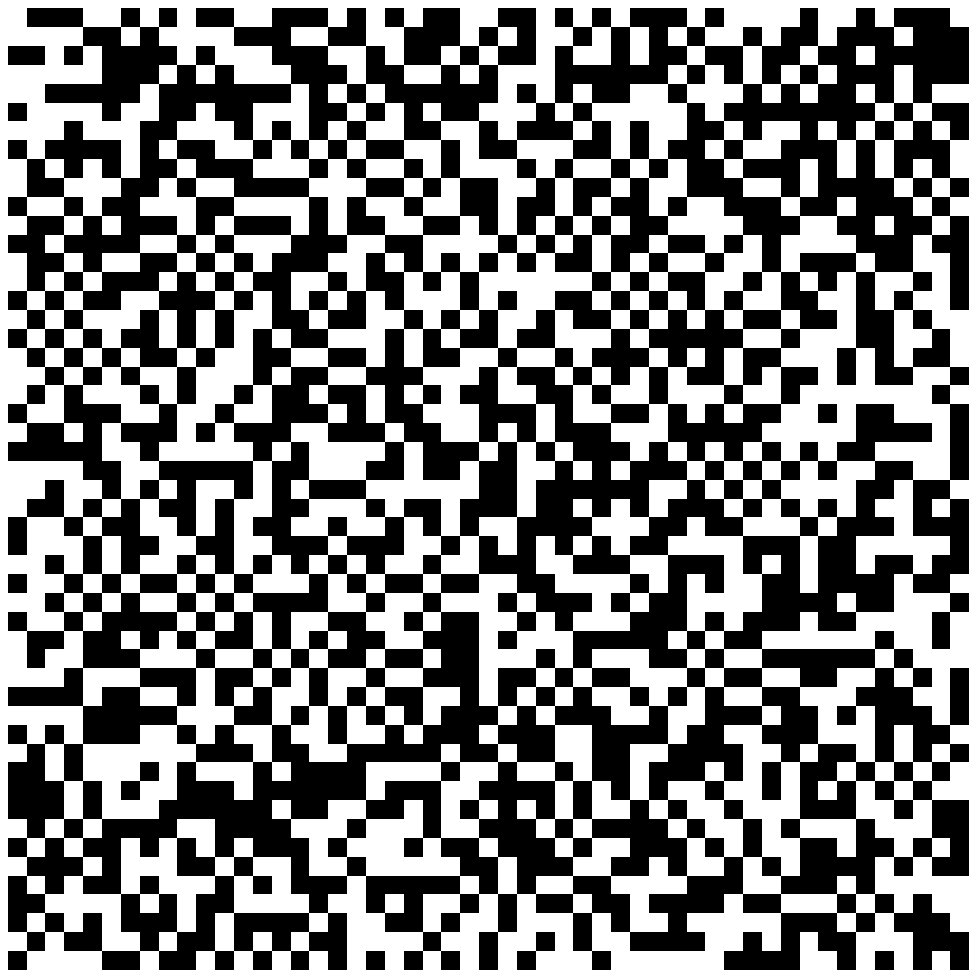}
&
\includegraphics[width=4.00cm,angle=0,clip]{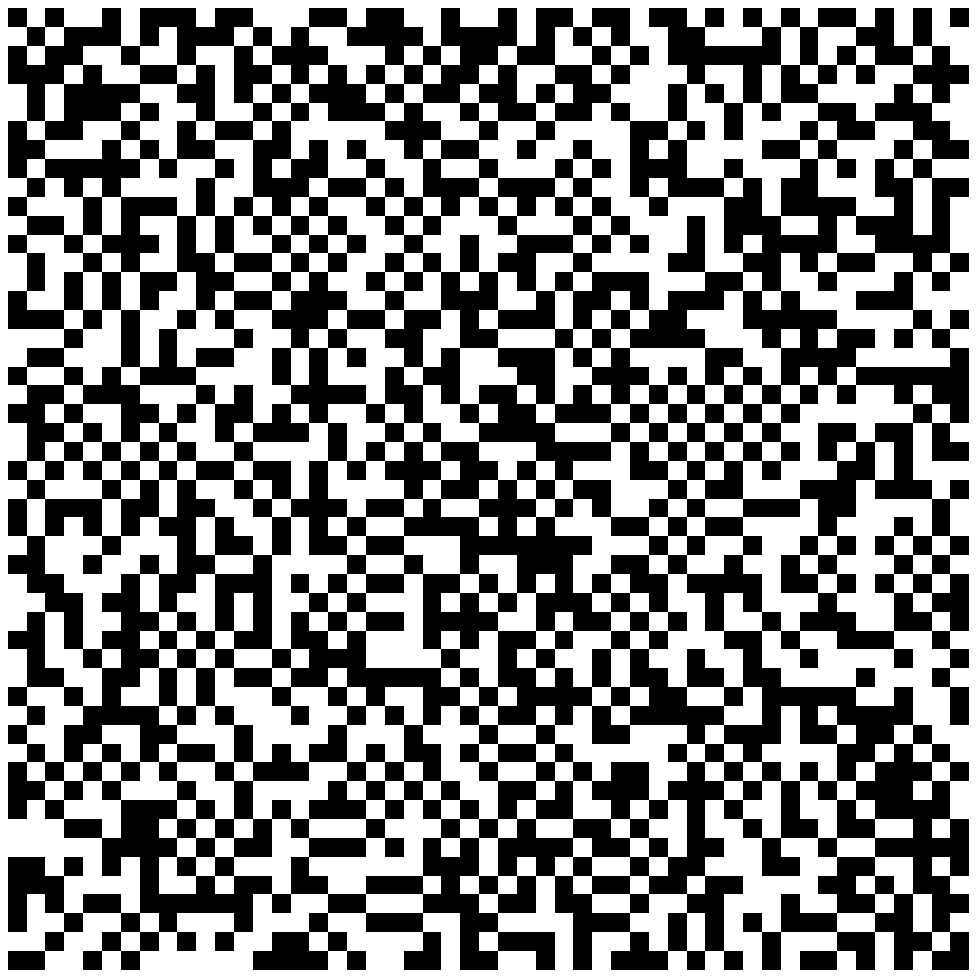}
\end{tabular}
\caption[a] { {\small The pattern of spin distribution at $T=1$
(a,b) and $T=3.5$ (c,d) without spin-lattice coupling for the
AFM case. Simulation results with 10$^3$ (a,c) and 10$^4$ (b,d) flips
per spin are presented. Black and white regions indicate the spin-up
and spin-down, respectively. } } \label{nn-spin}
\end{figure*}

\begin{figure*}[!htb]
\begin{tabular}{cccc}
 {\bf a} & {\bf b} & {\bf c} & {\bf d} \\
\includegraphics[width=4.00cm,angle=0,clip]{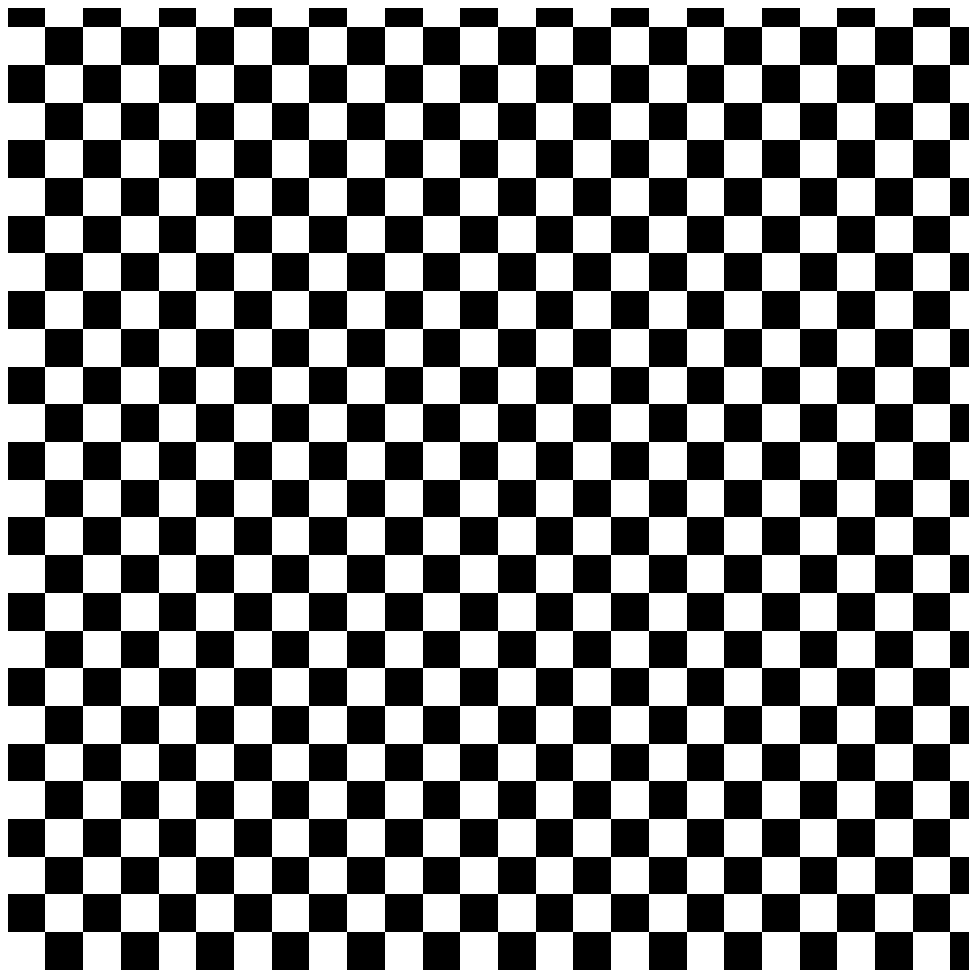}
&
\includegraphics[width=4.00cm,angle=0,clip]{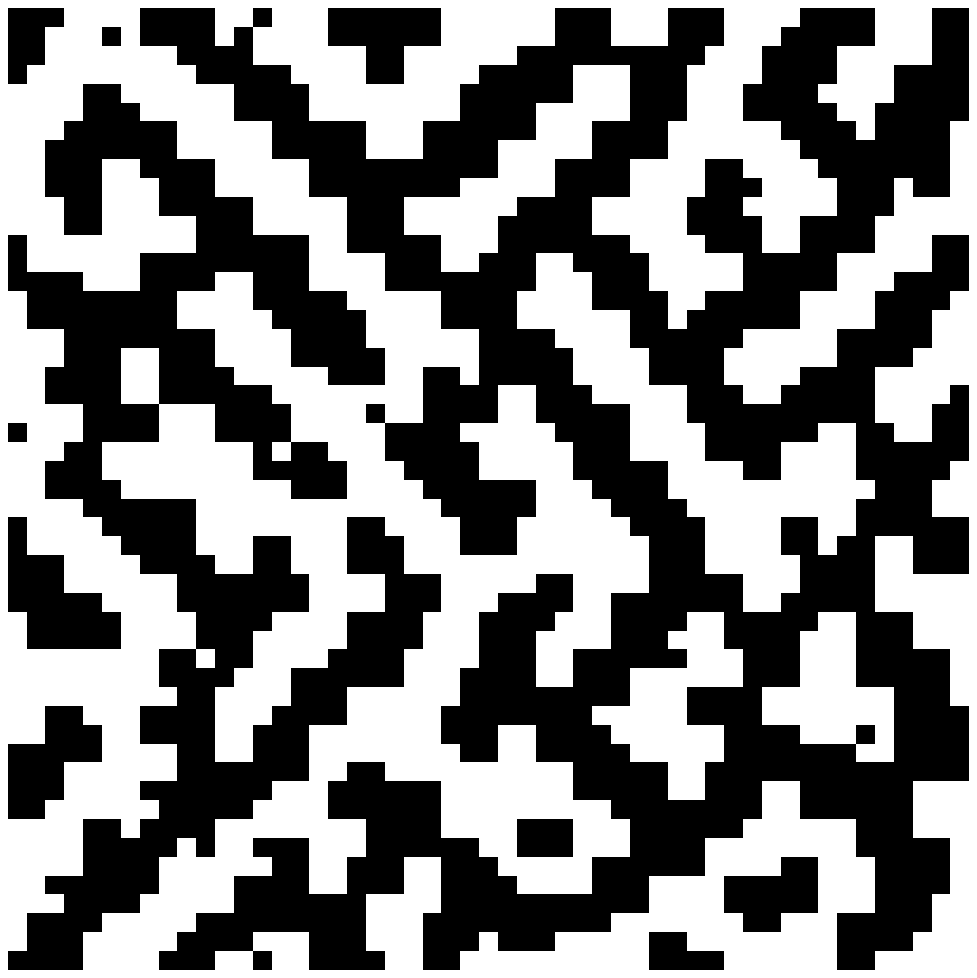}
&
\includegraphics[width=4.00cm,angle=0,clip]{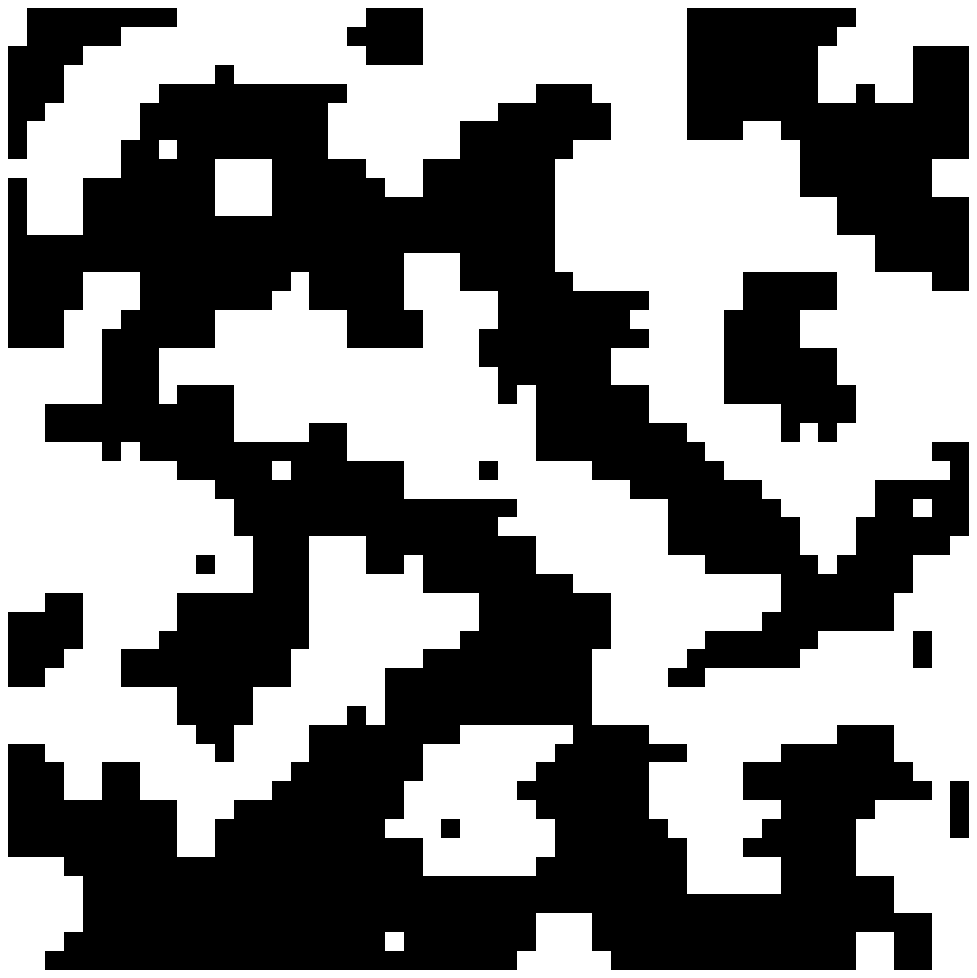}
&
\includegraphics[width=4.00cm,angle=0,clip]{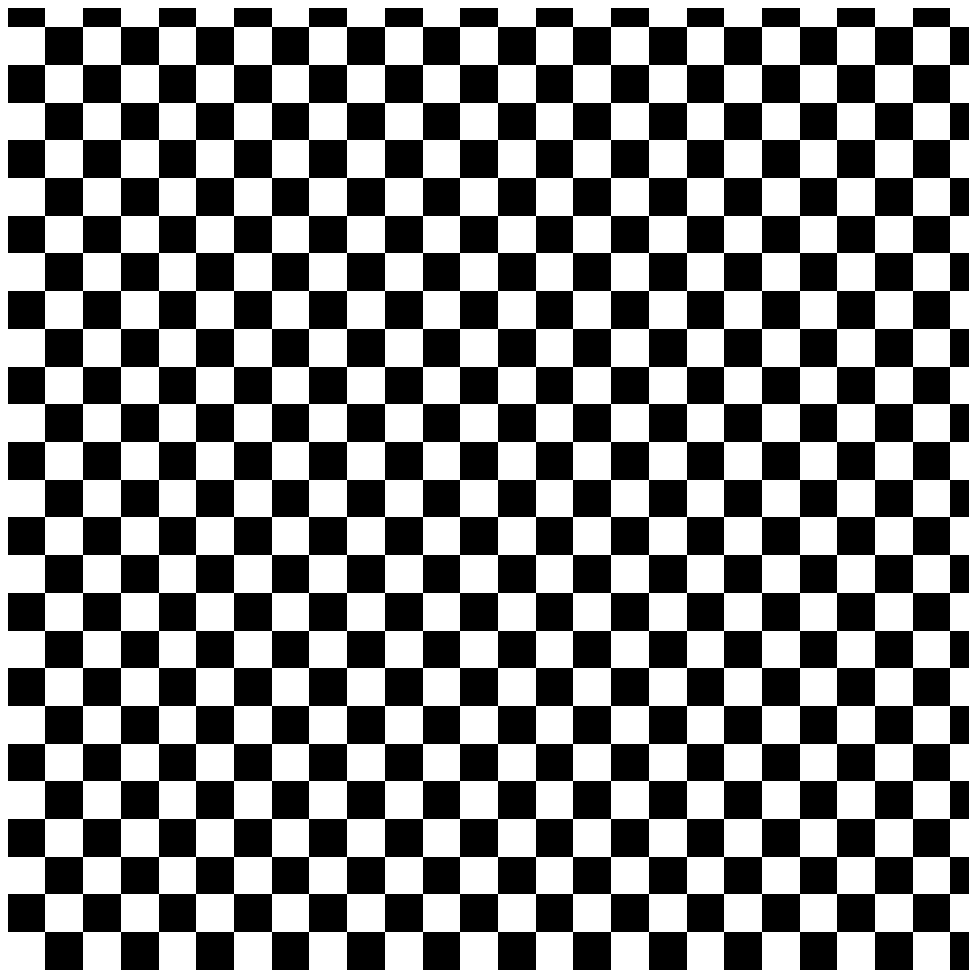}
\end{tabular}
\caption[a] { {\small The pattern of spin distribution at $T=1.5$
(a,b,c) and $T=3.5$ (d) with the spin-lattice coupling for
FM ground state at $Q=-10.0$ (a), $Q=-4.5$ (a), $Q=-2.5$ (a), $Q=-10.0$ (a).
} }
\label{lr-spin}
\end{figure*}

\section{Results of the Monte Carlo simulations}

In the square-lattice Ising model with {\it nn} interactions the
phase transition takes place at $T=T_c = 2.264$. One can see (Fig. \ref{nn-spin})
that the true ground state is reached for approximately 10$^4$ spin
flips per spin whereas for shorter Monte Carlo runs metastable configurations
such as domain walls appear. Further we will use by default this number of flips.
Probably it is not enough for a close vicinity of $T_c$ due to critical
slowing down but we will not discuss this region.

The effects of long-range interactions due to magnetoelastic
coupling are crucially dependent on dimensionless coupling
constant $Q=(J^{\prime})^2/\mu J^{0}_1$. Nontrivial patterns in a
broad enough temperature interval both below and above $T_c$ arise
at $|Q| > 3.5$; typical spin configurations
are shown in Fig. \ref{lr-spin}. In this case we have found a
long-range or short-range order of checkboard or stripe types with
staggered spin-up and spin-down regions. These patterns appear
when MC simulations start from both random and from regular
initial spin distribution and, therefore, are not metastable but,
rather, equilibrium states. Pictures of the long-range order are
quite similar for the case of FM ($J^{0}_1 = -1$) and AFM
($J^{0}_1 = 1$) Ising model.

\begin{figure}[!htb]
\begin{tabular}{cc}
{\bf a} & {\bf b}  \\
\includegraphics[width=3.95cm,angle=0,clip]{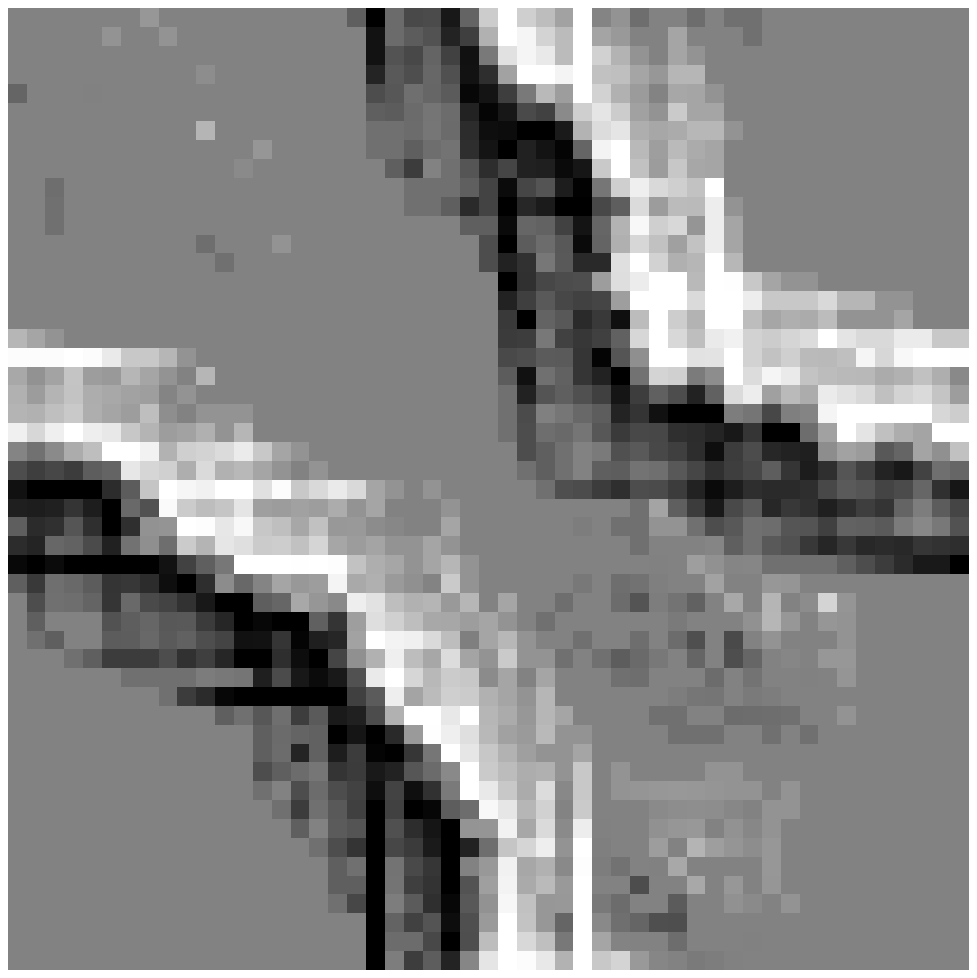}
&
\includegraphics[width=3.95cm,angle=0,clip]{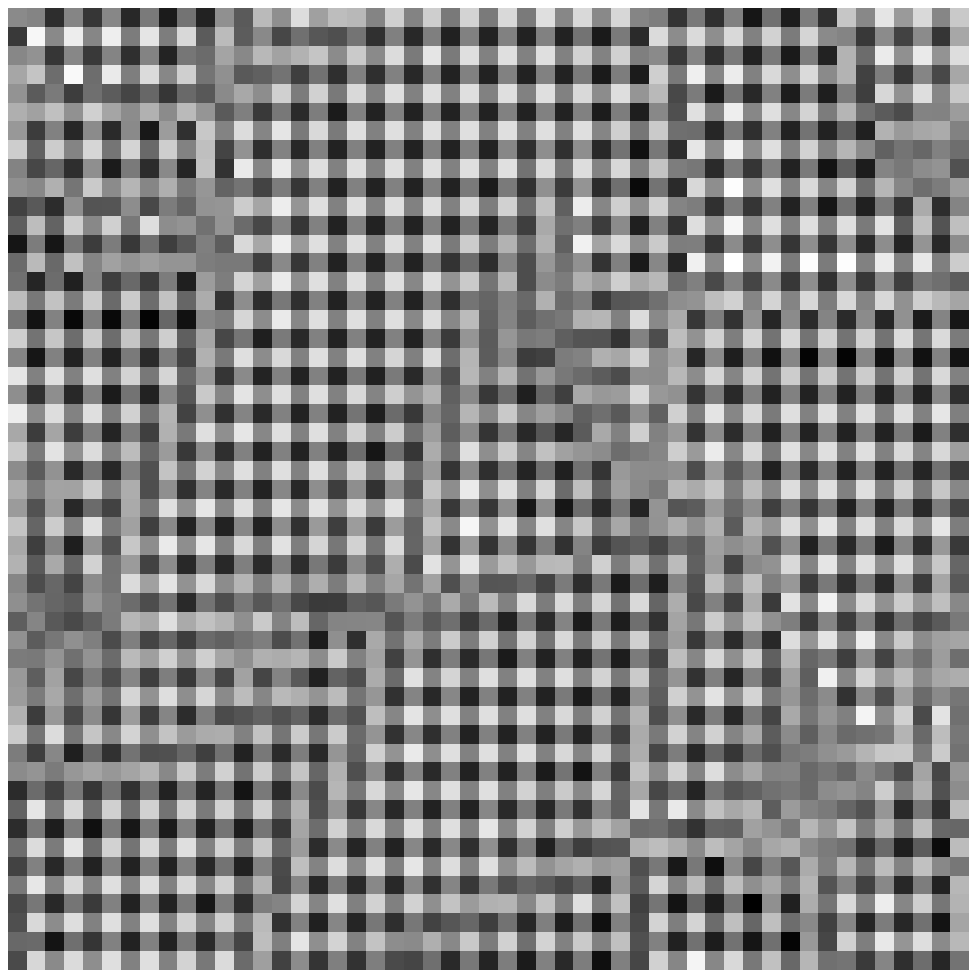}
\end{tabular}
\caption[a] { {\small Distribution of atomic displacements $u_x
+u_y$ for FM case at $T=1$ without spin lattice coupling ($Q=0$)
(a) and at $T=3.5$ with the spin-lattice coupling ($Q=-9.0$) (b).
Black regions correspond to $u_x +u_y<0$ and white ones to
$u_x+u_y>0$.} The results are presented for 10$^3$ flips per spin.}
\label{disp}
\end{figure}

Since these  new types of ordering result from the coupling between
magnetic and elastic subsystems it is naturally to expect its
manifestations also in the distribution of the displacement field
(Fig. \ref{disp}). For weak magnetoelastic coupling the lattice
distortions are created by domain boundaries in metastable configurations
and spread over large distances due to a slow decay of $G(r)$.
Fig. \ref{disp}a shows the corresponding distribution calculated by
the perturbation theory in the lowest order in $Q$. Spin patterns at large
$Q$ are inevitably related with the lattice distortion chessboard patterns
visible in Fig. \ref{disp}b. Contrary to previous case,
deformations are mutually cancelled at large spatial scale.

\begin{table}[h]
\caption{Energies (per atom) of different spin configurations; FM, AFM,
means the FM and AFM nearest-neighbour interaction; ``c''
$m$x$n$ labels checkboard ordering (see, e.g. Fig. \ref{lr-spin}b)
with elementary white and black $m$x$n$ rectangles;
``s'' $m$x$n$ labels diogonal stripes
with elementary steps $m$ and $n$ in $x$ and $y$ directions; ``R'' labels
random spin configuration. }
\label{tab:energy}
\begin{tabular}{ccccc} \\ \hline
 & \multicolumn{2}{c}{FM}  & \multicolumn{2}{c}{AFM}   \\ \hline
$Q$    &  -4 & -3 &  4 & 3    \\ \hline
FM     &  -4.0 & -4.0  &   4.0 & 4.0   \\
AFM    &   4.0 &  4.0  &  -4.0 & -4.0  \\
R      &  -1.8 & -1.0  &  -1.8 & -1.0    \\
c2x2   &  -5.3 & -3.0  &  -5.3 & -3.0   \\
c3x3   &  -4.2 & -2.9  &  -1.6 & -0.3   \\
c2x3   &  -4.8 & -3.0  &  -3.5 & -1.7   \\
s2x1   &  -4.7 & -2.6  &  -4.7 & -2.6   \\
s3x1   &  -4.2 & -2.9  &  -1.5 & -0.3   \\
\hline
\end{tabular}
\end{table}

\begin{figure}[!htb]
\includegraphics[width=7.6cm,angle=0,clip]{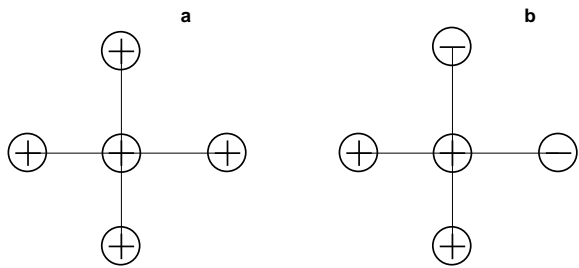}
\caption[a] { {\small Local spin configurations forming different patterns.
 } }
\label{sketch}
\end{figure}

\begin{figure}[!htb]
\includegraphics[width=7.9cm,angle=0,clip]{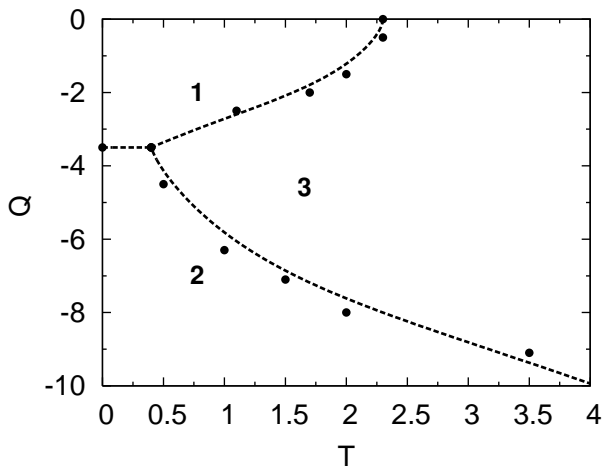}
\caption[a] { {\small A phase diagram  for ferromagnetic
case.
1, 2, and 3 labels FM phase, checkboard ordering, and disordered
(paramagnetic) phase, respectively.
 } }
\label{phase}
\end{figure}

To understand character of new types of ordering let us consider
energies of different local spin configurations (Fig.
\ref{sketch}). For ferromagnetic ordering (Fig. \ref{sketch}a) the
force (\ref{P}) $P_i ^{\alpha} = 0$ and therefore the
magnetoelastic contribution to the total energy vanishes. On the
other hand, the configuration Fig. \ref{sketch}b corresponds to
the largest local value of $P_i ^{\alpha}$ and, thus, to the
maximal magnetoelastic energy gain. This configuration is a
minimal structural block of stripe and checkboard structures.
Direct total energy calculations for various spin configurations
(see the Table) demonstrate that at $Q \approx -3.5$ simple FM or
AFM structure become energetically unfavorable and the new ordered
ground state occurs. In this case the checkboard ordering is
preferable and the stripe structure has slightly higher energy.

With the temperature increase, the long-range order is destroyed.
We estimate critical temperatures of these transitions calculating
the $R$-dependence of pair correlation functions of the corresponding
order parameters.
The results are presented in Figure \ref{phase}. With the growth of
$Q$ the temperature of transition from FM (AFM) and paramagnetic phase
decreases. This is not surprizing since the magnetoelastic coupling
destabilizes these spin states. With further increase of $Q$ there is
a transition between check-board (or stripe) phase and the paramagnetic
state, the transition temperature growing with $Q$.

\begin{figure}[!htb]
\includegraphics[width=7.4cm,angle=0,clip]{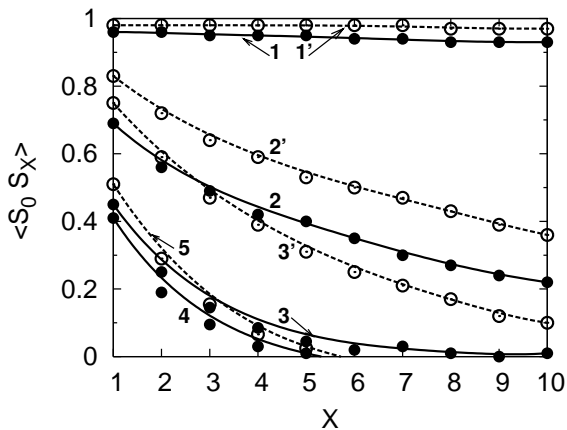}
\caption[a] { {\small Space dependence of the correlation function
$<S_0S_X>$ in $X$ (or $Y$) direction for $Q=0$ (solid curves 1, 2,
3, and 4) and $Q= - 2.0$ (dashed curves 1', 2', 3', and 5). The
temperatures are equal to 0.9 $T_c$ (curves 1, 1'), 1.1 $T_c$
(curves 2, 2'), 1.2 $T_c$ (curves 3, 3'), 1.3 $T_c$ (curve 4), and
1.65 $T_c$ (curve 5).
 } }
\label{corr}
\end{figure}

Figure \ref{corr} shows the evolution of the spin correlation
functions for $|Q| < 3.5$ (that is, at the transition between
regions 1 and 3 at the phase diagram \ref{phase}) with the
temperature increase. One can see that the magnetoelastic coupling
leads to stronger short-range order above $T_c$ (c.f. curves 2 and
2', 3 and 3', respectively), which survives up to a relatively
higher temperature, in comparison with the original Ising model.
In particular, the short-range order for $Q=-2.0$ and $T=1.65T_c$
(curve 5) is stronger than for $Q=0$ and $T=1.3T_c$ (curve 4).

The short-range order in the region 3 near the triple point tends
to formation of stripe-like structures (Fig. \ref{lr-spin}b,c),
although this configuration is not the most energetically
preferable. The appearance of this structure is connected with its
higher entropy in comparison with the checkboard structure.

\section{Discussion and conclusions}

Our computational results are summarized schematically in the
phase diagram shown in Fig. \ref{phase}. For small enough coupling
constant $|Q| \leq 3.5$ the behavior of the system is qualitatively
similar to that of the standard Ising model. In this case the main
effect of the magnetoelastic coupling is suppression of $T_c$ due to
destabilization of FM or AFM states (Fig. \ref{corr}). Usually this
happens if $T_c$ is strongly suppressed by soft fluctuations, such
as in quasi-two-dimensional Heisenberg magnets~\cite{SSWT1} or
frustrated three-dimensional Heisenberg magnets with competing
interaction~\cite{SSWT2}. In those cases $T_c$ is much smaller
than a typical energy of nearest-neighbor interaction $J$ whereas
the short-range order in a broad temperature interval survives up
to the temperatures of order of $J$. We illustrate here an
alternative mechanism of occurrence of the short-range order due
to coupling with a second subsystem providing long-range effective
interactions.

Thus, in this regime the magnetoelastic interactions are favourable for
the short-range order in paramagnetic phase but do not result yet in the
nontrivial pattern formation.
The latter happens at $Q>3.5$ where exotic spin configurations
(checkboard or stripe states) become
favourable. Especially,
near the triple point at the phase diagram (Fig. \ref{phase}) there are many
different states with approximately the same free energy and therefore a
formation of complicated inhomogeneous states could be expected. At further
temperature increase we reach a conventional paramagnetic (random)
state without short-range order. This scenario reminds the behavior
observed at the ``melting'' of stripe domains in magnetic films~\cite{debell},
our ``pattern'' regime being similar to ``tetragonal phase'' in the latter case.
The main formal difference between these two problems is two-spin
character of long-range dipole-dipole interactions versus
four-spin character of long-range phonon-mediated interaction
considered here.

Despite a simplicity of our model, it demonstrates a rather
general feature which we believe can be relevant for discussions
of real systems demonstrating a strong short-range order above
$T_c$ in a broad temperature interval. The effect arises at large
enough coupling constant $|Q|$ which means either unusually strong
interaction between the subsystems (large $J'$) or soft phonons
(small shear modulus). At least, in some case, such as Cu-Mn
alloys the shear modulus tends to zero at some critical composition
or temperature~\cite{tsunoda} so $|Q|$ can be, in principle, arbitrarily
large. It would be interesting to analyse the problem of heterophase
fluctuations in alloys mentioned in the Introduction from this
point of view.

\textbf{Acknowledgements.} The work was supported by the
Netherlands Organization for Scientific Research (NWO project
047.016.005) and by the Stichting Fundamenteel Onderzoek der
Materie (FOM).

\end{document}